\input amstex
\input amsppt.sty
 \pageheight{18.5cm}
\magnification=\magstep1
\TagsOnRight
\NoBlackBoxes
\NoRunningHeads
\nologo
\topmatter
\title Blowup of smooth solutions for relativistic Euler equations 
\endtitle
\author  Ronghua Pan$^{1}$ and Joel A. Smoller$^{2}$
\endauthor
\affil 
 1. School of Mathematics, Georgia Institute of Technology.\\
 2. Department of Mathematics, University of Michigan .
\endaffil
\address 
 \endaddress
\email  panrh\@math.gatech.edu, smoller\@umich.edu
\endemail
\abstract We study the singularity formation of 
smooth solutions of the relativistic Euler equations in $(3+1)$-dimensional
spacetime for both finite initial energy and infinite initial energy.
For the finite initial energy case, we prove that any smooth solution,
with compactly supported non-trivial initial data, blows up  in
finite time.   
For the case of infinite initial energy, we first prove the 
existence, uniqueness and stability of a smooth solution if the
initial data is in the subluminal region away from the vacuum.
By further assuming the initial data is a smooth compactly supported 
perturbation around a non-vacuum constant background, we prove the property of
 finite propagation speed of such a perturbation. The smooth solution is
 shown to blow up in finite time provided that
the radial component of the initial ``generalized" momentum is sufficiently large. 
\endabstract
\endtopmatter
\document

\subheading{1. Introduction}\par

In this paper, we study the singularity formation of solutions of the
Einstein equations for an isentropic perfect fluid. Due to the hyperbolic
nature of these nonlinear equations, one expects singularity formation in
the solutions. Indeed, one even expects black holes to form. 
However,  singularity formation in relativistic 
flow is not yet well-understood;  the theory is 
most lacking in the multi-dimensional case,
 $(3+1)$-dimensional spacetime.
 
As a first step in this direction, we consider here the 
relativistic Euler equations for a perfect
fluid in 4-dimensional Minkowski spacetime,    

 $$ Div \ T=0,\tag 1.1$$
where

$$T^{ij}=(p+\rho c^2){\bold u}^{i}{\bold u}^{j}+pg^{ij},\tag 1.2$$
is the stress-energy tensor for a perfect fluid, and $g^{ij}$ denotes
the flat Minkowski metric, $g^{ij}=diag(-1, 1,1,1)$, 
${\bold x}=(x^0, x^1, x^2, x^3)^T$ with $x^0=ct$. $\rho$ is the
mass-energy density, $p$ is the pressure, $c$ is the speed of light,
and ${\bold u}$ is the $4$-velocity of the fluid. Recall that since 
$\displaystyle {\bold u}=\frac{1}{c}
\frac{d{\bold x}}{d\tau}$ ($\tau$ is the proper time, ${\bold u}$ is a unit 
$4$-vector in Minkowski space), it follows that 

   $$(u^0)^2-\sum_{\alpha=1}^{3} (u^\alpha)^2=1,$$
and thus only three of the quantities $u^0$, $u^1$, 
$u^2$, $u^3$ are independent. We now fix our space-time coordinates
as $(t, x^1, x^2, x^3)^T$, set $x=(x^1, x^2, x^3)^T$,
 $u=(u^1, u^2, u^3)^T$, and let
$$v=\frac{cu}{\sqrt{(1+|u|^2)}}.$$
One easily derives from equation (1.1) the relativistic Euler equations:

 $$\cases
    &\displaystyle \partial_t(\frac{\rho c^2+p}{c^2-v^2}-\frac{p}{c^2})
         +\nabla_x\bullet (\frac{\rho c^2+p}{c^2-v^2} v)=0\\
      &\displaystyle \partial_t(\frac{\rho c^2+p}{c^2-v^2} v)
         +\nabla_x\bullet(\frac{\rho c^2+p}{c^2-v^2} v\otimes v)
           +\nabla_x\ p=0,\endcases\tag 1.3$$
in the unknowns $\rho$, $v$ and $p$. Here $\nabla_x$ denotes the 
spatial gradient operator. Given a scalar $k$ and 3-vectors ${\bold a}$ 
and ${\bold b}$, by the notion ${\bold a}\otimes {\bold b}$ we mean
the matrix ${\bold a}{\bold b}^T$, while 
$$\nabla_x\bullet(k{\bold a}{\bold b}^T)=(\nabla_x\bullet (ka_1{\bold b}),
 \nabla_x\bullet (ka_2{\bold b}),\nabla_x\bullet (ka_3{\bold b}))^T.$$                                          
 We consider the Cauchy problem for (1.3)
with initial data 
$$\rho(x,0)=\rho_0(x), v(x,0)=v_0(x).\tag1.4$$
Equations (1.3) close if we assume 
an equation of state, $p=p(\rho)$, $p(0)=0$ with
$$p(\rho)\ge 0,\ 0<p'(\rho)<c^2,\  p''(\rho)\ge 0,\ for \ \rho \in (\rho_*, \rho^*),
\tag1.5$$
where $0\le\rho_*<\rho^*\le\infty$. For a $\gamma$-law, 
$p(\rho)=\sigma^2\rho^\gamma$ with $\gamma\ge 1$,   
the constant $\rho^*$ is chosen as follows:
if $\gamma=1$, then $\rho^*=\infty$; and if 
$\gamma>1$, then $p'(\rho^*)=c^2$. 
Thus the unknowns for the Cauchy problem (1.3)--(1.4) are 
$\rho$ and $v$. For more details on the derivation of equations (1.3) 
and a discussion of (1.5), see [14].

We are interested in the life span of smooth solutions for the Cauchy
problem (1.3)--(1.4). For this purpose, we shall discuss two different
cases: the case of finite initial energy, and the case
of infinite initial energy. For the first case, we shall prove
that  if the initial data has compact support,  then 
the life span of any non-trivial smooth solution
for the Cauchy problem (1.3) and (1.4) is finite. 
For the second case,  we show that if the initial 
data is a compactly supported 
perturbation around a non-vacuum background, then the life span of smooth
solutions is finite provided that the radial component of the initial 
``generalized" momentum is sufficiently large; c.f. Theorem 3.2.

We start with the infinite energy case.
The local existence of classical solutions of the Cauchy problem (1.3)--(1.4) 
has been established by Makino and Ukai ([6, 7])
provided that the initial data is in the subluminal region away from the
vacuum. A sharper result is proved here in Theorem 2.1 in Section 2, 
where the stability of the solution with respect to the initial data 
(c.f. Corollary 2.2) and the
properties of finite propagation speed (c.f. Lemma 2.3) are 
presented. In Section 3, we first derive some interesting structural 
properties of (1.3) in Lemma 3.1, then we prove a blowup result
for smooth solutions
(c.f. Theorem 3.2). Our proof
is in the spirit of the work of Sideris for classical
Euler equations [11] and is based on the largeness of 
the initial radial component of ``generalized" momentum,
which of course implies the largeness of the initial velocity. However,
in our case, the velocity is still subluminal.  
In Section 4, we prove our blowup result for smooth solutions of
(1.3)--(1.4) with non-trivial initial data that has compact support.
In Section 5, we make some remarks concerning
our results. A discussion on the type of singularity 
is also given. The existence of initial data satisfying our blowup
conditions is also shown there.  All the results in Section 2 
are based on the existence of a strictly convex entropy function for (1.3), 
which was constructed by Makino and Ukai
in [6, 7].  For the reader's convenience, 
we present the construction in the Appendix, 
correcting a few errors in the original papers. \par 

Before proceeding,
we now briefly review the methods and results of singularity formation
for nonlinear hyperbolic systems. In one space dimension, the theory
is fairly complete. It was proved that 
a singularity develops in finite time no matter how
small and smooth the initial data is; c.f. [4, 5, 13].
These results were established by the characteristic method, 
which is quite powerful
in one space dimension. In more than one space dimension, there are
no general theorems available mainly because the characteristics
become intractable. However, the approach via certain averaged quantities was
introduced by Sideris [11] to prove the formation of singularities
in three-dimensional compressible fluids. This idea avoids the local
analysis of solutions. A similar technique was used to prove other
formation of singularity theorems. We refer to [8], [9],
[16] for classical fluids, and [2], [10] for relativistic 
fluids. Blowup results for relativistic Euler equations are announced in
[2] and [10]. However, as remarked on page 154 of [2],
`` the unpublished proof in [10] contained an error which 
invalidated the argument ". Furthermore, we note that the coefficient 
matrices in (2.15) of [2] 
constructed through (2.16) of [2] 
are not symmetric away from the equilibrium. But
the symmetry of (2.15) in [2] is crucial to prove the finite propagation speed 
property needed in their proof. Thus the argument in [2] is not complete.
Furthermore,  we note that the equation of state used in [2] and [10]
is different from ours.  In addition, the approach of [2] is also 
different from ours.
Our approach is closer to the method of Sideris [11]. Finally, we remark that
the equation of state (1.5) in this paper is interesting for cosmology. 
It includes many physical cases, e.g. $\gamma$-laws, 
$p(\rho)=\sigma^2\rho^\gamma$, $\gamma\ge1$.  For instance, the case 
$$ p(\rho)=\frac13 c^2 \rho$$ 
is very important in cosmology; 
it is the equation of state for the Universe in earliest times
after the Big-Bang; see [15]. Some cases discussed in [2] (e.g. 
when $s=const.$) satisfy (1.5) as well.
Another important example (see [15, p.319]) is the equation of state for
neutron stars, where 

$$\aligned& p=Ac^5a(y), \ \rho=Ac^3b(y),\\
 &a(y)=\int_0^y \frac{q^4}{\sqrt{1+q^2}}\ dq,\ 
 \ b(y)=3\int_0^y q^2\sqrt{1+q^2}\ dq.
 \endaligned\tag 1.6$$
Here $A$ is a positive constant. This equation of state implies
the following asymptotics: 
$p\to \frac13 c^2 \rho$ as $\rho\to \infty$
and $p\to \frac15 A^{2/3}\rho^{5/3}$ as $\rho\to0$. It is easy to see 
that 
$$ p'(\rho)=\frac{c^2y^2}{3(1+y^2)}>0,
\ p''(\rho)=\frac{2}{9Acy} (1+y^2)^{-5/2}>0,$$
whenever, $y>0$. We also note that $y=0$ is equivalent to $\rho=0$. 
Thus the equations (1.6) also satisfy (1.5).\par

\subheading{2. Existence of Solutions: Infinite Energy Case}

In this section, we consider the local existence of smooth solutions for
the Cauchy problem (1.3)--(1.4) when the initial
data is away from the vacuum. For this purpose, 
we introduce some convenient notation:

$$\aligned & \displaystyle {\tilde\rho}=   \frac{\rho c^2+p}{c^2-v^2},\\
    & \displaystyle 
 {\hat\rho}=(\frac{\rho c^2+p}{c^2-v^2}-\frac{p}{c^2}).
\endaligned\tag2.1$$
The Cauchy problem (1.3)-(1.4)  becomes

$$\cases
      &{\hat \rho}_t+\nabla_x\bullet({\tilde\rho}v)=0\\
      &({\tilde \rho}v)_t+\nabla_x\bullet ({\tilde \rho}v\otimes v)+
       \nabla_x p(\rho)=0,\\
       &{\rho}(x, 0)={ \rho}_0(x),\ \ v(x, 0)=v_0(x).
      \endcases\tag2.2$$

Let $\rho_*< \rho^*$ be non-negative constants in (1.5) subject to
the subluminal condition $p'(\rho^*)\le c^2$. We set 
$$z=(\rho, v_1, v_2, v_3)^T$$
 and define the region $\Omega_z$ by
$$\Omega_z=\{z: \rho_*<\rho<\rho^*, v^2<c^2\}.\tag 2.3$$

\proclaim{Theorem 2.1} Assume an equation of state is given as in (1.5).
Suppose the initial data $z_0(x)=(\rho_0(x), v_0(x))^T$ is continuously 
differentiable on ${\bold R}^3$, taking values in any compact subset
${\bold D}$ of $\Omega_z$ and that
$\nabla_x z_0 (x)\in H^l({\bold R}^3)$ for some $l>3/2$. Then there exists
$T_\infty$, $0<T_{\infty}\le \infty$, and a unique differentiable function
$z(x,t)=(\rho(x,t), v(x,t))^T$ on ${\bold R}^3\times [0, T_\infty)$, taking
values in $\Omega_z$, which is a classical solution of the Cauchy problem 
(1.3)--(1.4) on ${\bold R}^3\times [0, T_\infty)$. Furthermore,

  $$\nabla_x z(\cdot, t)\in C^0([0, T_\infty); H^l). \tag 2.4$$
The interval $[0, T_\infty)$ is maximal, in the sense that whenever
$T_\infty<\infty$,
$$\underset{t\to T_\infty}\to{\lim}\sup \|\nabla_x z(\cdot, t)\|_{L^\infty}
     =\infty\tag2.5$$
and/or the range of $z(\cdot, t)$ escapes from every compact subset of 
$\Omega_z$ as $t\to T_{\infty}$. 
 
\endproclaim\par

This theorem will be proved by applying Theorem 5.1.1 in
Dafermos [1] for hyperbolic conservation laws endowed with a strictly
convex entropy. We state this theorem here for readers convenience. 

\proclaim{Theorem A} Assume that the system of conservation laws 
     $$U_t+\sum_{\alpha=1}^m \partial_{x_\alpha}G_\alpha (U)=0,
                   \ x\in {\bold R}^m, U\in {\bold O}
\subset{\bold R}^n,\tag *$$
is endowed with an entropy $\eta$ with $\nabla^2\eta(U)$ positive 
definite, uniformly on a compact subset of ${\bold O}$. Suppose the 
initial data $U(x,0)=U_0(x)$ is continuously differentiable on 
${\bold R}^m$, takes values in some compact subset of ${\bold O}$
and $\nabla U_0\in H^l$ for some $l>m/2$. Then there exists $T_\infty$,
$0<T_\infty\le \infty$, and a unique continuously differentiable 
function $U$ on ${\bold R}^m\times[0,T_\infty)$, taking values in 
${\bold O}$, which is a classical solution of the initial-value
problem $(*)$ with initial data $U_0$ on $[0,T_\infty)$. Furthermore,
    $$\nabla U(\cdot, t)\in C^0([0, T_\infty); H^l).$$
The interval $[0, T_\infty)$ is maximal, in the sense that whenever 
$T_\infty<\infty$

$$\underset{t\to T_\infty}\to{\lim}\sup\|\nabla U(\cdot, t)\|_{L^\infty}
    =\infty$$
and/or the range of $U(\cdot,t)$ escapes from every compact subset of
${\bold O}$ as $t\to T_\infty.$
\endproclaim\par
 
\demo{Proof of Theorem 2.1} We first rewrite (1.3) or (2.2) in the 
form of  conservation laws,
 
 $$\theta_t+\sum_{k=1}^3 (f^k(\theta))_{x_k}=0,\tag2.6$$
 where $\theta=(\theta_0, \theta_1, \theta_2,\theta_3)^T$ and 
 $f^k(\theta)=(\theta_k, f_1^k, f_2^k, f_3^k)^T$
 are defined by
 
 $$\aligned & \theta_0={\hat \rho},\ \theta_j={\tilde\rho}v_j,\\
    &f_j^k={\tilde\rho}v_jv_k+p\delta_{jk},\ j=1,2,3.\endaligned\tag2.7$$
By Theorem A, it is sufficient to show that (2.6) has an entropy
$\eta(\theta)$ with $\nabla^2\eta(\theta)$ positive definite in $\Omega_z$.
Such an entropy, due to Makino and Ukai [7], is constructed in 
the Appendix of this paper.\par 

Define 
$$\phi(\rho)=\int_{\rho_m}^\rho \frac{c^2}{r c^2+p(r)}\ dr,
\  \  K={\rho_m}c^2+p(\rho_m),
\tag 2.8$$
 ${\rho_m}$ being any fixed number in $(\rho_*, \rho^*)$. 
The entropy given in (6.25) below is 
 
 $$\eta=c^2{\hat\rho}-  \frac{cKe^{\phi(\rho)}}{\sqrt{c^2-v^2}}.\tag 2.9$$
We now verify that $\nabla^2\eta (\theta)$ is positive definite 
in $\Omega_z$.  To this end, we first compute $\nabla_{\theta}\eta(\theta)$. 
By the chain rule, we have
$$w^T=(\nabla_{\theta}\eta)=(\nabla_z \eta)(\nabla_z \theta)^{-1},$$
 where  $(\nabla_z \theta)^{-1}$ is defined in (6.10), and
  $ w^T=(w_0, w_1, w_2, w_3)$ is given by
$$\cases &w_0=-\frac{c^3\Phi(\rho)}{(c^2-v^2)^{1/2}} +c^2,\\
  &w_j=\frac{c\Phi(\rho)}{(c^2-v^2)^{1/2}} v_j,\ j=1,2,3, \endcases\tag2.10$$
  with
  $$\Phi(\rho)=\frac{Ke^{\phi(\rho)}}{(\rho c^2+p)}.\tag 2.11$$
We remark that $w$ can serve as a symmetric variable which reduces 
(1.3) to a symmetric hyperbolic system [1, 3]. 
For the Hessian matrix $H$ of $\eta$,
we compute 
$$\aligned H=&  \nabla^2\eta (\theta) 
=\nabla_{\theta} w^T=(\nabla_z w^T)(\nabla_z \theta)^{-1}\\
  &=\frac{c\Phi(\rho) E_1}{(\rho c^2+p)(c^2-v^2)^{1/2}}H_1\\
 &\equiv \frac{c\Phi(\rho) E_1}{(\rho c^2+p)(c^2-v^2)^{1/2}} 
\left( \matrix A_1&A_2v^T\\
          A_2v& A_3vv^T+A_4I_3 \endmatrix\right).\endaligned\tag 2.12$$
Here, $E_1=\frac{1}{c^4-p'v^2}$ is given in (6.11) below, and the $A_i$ are  given by
$$\aligned & A_1=c^4(p'c^2+2p'v^2+c^2v^2),\ \ A_2=-c^2(c^4+2c^2p'+p'v^2),\\
   &A_3=(c^4+2c^2p'+p'v^2+2p'(c^2-v^2)),\ \ 
   A_4=(c^2-v^2)(c^4-p'v^2).\endaligned\tag2.13$$
   
We now show that $H$ is positive definite. From (2.12), we see that it 
 is sufficient to show $H_1$ is positive definite.  
 Let ${\bold r}=(r_0, r^T)^T$ be any 4-vector with
 $r\in {\bold R}^3$. We calculate:
 $$\aligned {\bold r}^TH_1 {\bold r}&
=(r_0, r) \left( \matrix A_1&A_2v^T\\
          A_2v& A_3vv^T+A_4I_3 \endmatrix\right)(r_0, r)^T \\
        &=(A_1r_0^2+ 2A_2r_0v^Tr+A_3(v^Tr)^2+A_4r^2).\endaligned$$
Letting ${\tilde A}_1=(1-\delta)A_1$ with $\frac12>\delta>0$ to be 
determined in (2.14) below, we have 

$$\aligned& (A_1r_0^2+ 2A_2r_0v^Tr+A_3(v^Tr)^2+A_4r^2)\\
     &={\tilde A}_1(r_0+\frac{A_2}{{\tilde A}_1} v^Tr)^2
-(\frac{1}{A_1}(A_2^2-A_1A_3)
+\frac{\delta}{1-\delta} \frac{A_2^2}{A_1})(v^Tr)^2+\delta A_1r_0^2+A_4r^2\\
&\ge 
(A_4-\frac{1}{A_1}(A_2^2-A_1A_3)v^2
-\frac{\delta}{1-\delta} \frac{A_2^2}{A_1}v^2)r^2+\delta A_1r_0^2\\
     &\ge (\frac{p'(c^2-v^2)^2(c^4-p'v^2)}{(p'c^2+2p'v^2+c^2v^2)}
     -2\delta \frac{A_2^2}{A_1}v^2)r^2+\delta A_1r_0^2\\
    &\ge  \delta A_1r_0^2+\delta  r^2.\endaligned$$
Here, we determine $\delta$ by

$$0<\delta+2\delta  \frac{A_2^2}{A_1}v^2
< \frac{p'(c^2-v^2)^2(c^4-p'v^2)}{(p'c^2+2p'v^2+c^2v^2)}.\tag2.14$$
We thus conclude that

$${\bold r}^TH_1 {\bold r}\ge  
(\delta A_1r_0^2+\delta  r^2).$$ 
This proves $H_1$ is positive definite in $\Omega_z$. 
Hence, $H$ is positive definite,
and $\eta$ is strictly convex on $\Omega_z$. This
completes the proof of Theorem 2.1. 
\qquad\qed\enddemo\par

\par

The existence of a strictly convex entropy guarantees that 
classical solutions of the initial-value problem depend continuously
on the initial data, even within the broader class of admissible
bounded weak solutions; see [1].  Here, by admissible bounded weak solution, we
mean bounded functions satisfying the initial value problem and entropy 
inequality in the
sense of distributions. The following Theorem B is 
Theorem 5.2.1 in Dafermos [1]:

\proclaim{Theorem B} Assume that the system of conservation laws 
$(*)$ is endowed
with an entropy $\eta$ with $\nabla^2 \eta(U)$ positive definite, uniformly
on compact subset of ${\bold O}$. Suppose $U$ is a classical solution of
$(*)$ on $[0, T)$, taking values in a convex compact subset $N$ of ${\bold O}$,
with initial data $U_0$. Let ${\bar U}$ be any admissible weak solution of 
$(*)$ on $[0, T)$, taking values in $N$, with initial data ${\bar U}_0$. Then

 $$\int_{|x|<R}|U(x,t)-{\bar U}(x,t)|^2\ dx
   \le a e^{bt}\int_{|x|<R+st}|U_0(x)-{\bar U}_0(x)|^2 \ dx$$
holds for any $R>0$ and $t\in [0, T)$, with positive constants
$s$, $a$, depending only on $N$, and a constant $b$ that also depends
on the Lipschitz constant of $U$. In particular, ${\bar U}$ is
the unique admissible weak solution of $(*)$ with initial data ${\bar U}_0(x)$
and values in ${N}$.
\endproclaim\par

The following corollary is a consequence
of the convexity of $\eta$ and Theorem B.

\proclaim{Corollary 2.2} Let $\theta$ be a classical solution of (2.6)
obtained in Theorem 2.1 with initial data $\theta_0(x)$ taking values in a
compact subset ${\bold D}$ of $\Omega_z$, and let ${\tilde \theta}$ be any admissible
weak solution of (2.6) on $[0, T_\infty)$, taking values in ${\bold D}$  , with initial value 
${\tilde \theta}_0(x)\in {\bold D}$. Then 

$$\int_{|x|<R}|\theta(x,t)-{\tilde\theta}(x,t)|^2\ dx
   \le a e^{bt}\int_{|x|<R+st}|\theta_0(x)-{\tilde \theta}_0(x)|^2 \ dx$$
holds for any $R>0$ and $t\in [0, T_\infty)$, with positive constants
$s$, $a$, depending only on ${\bold D}$, and a constant $b$ that also depends
on the Lipschitz constant of $\theta$. In particular, ${\tilde\theta}$ is
the unique admissible weak solution of (2.6) with initial data 
${\tilde\theta}_0(x)$ and values in ${\bold D}$.
\endproclaim\par

In the next lemma, we will show that a compactly supported perturbation
around a non-vacuum background propagates with finite speed. For this 
purpose, we consider the following Cauchy problem

$$\cases
      &{\hat \rho}_t+\nabla_x\bullet({\tilde\rho}v)=0\\
      &({\tilde \rho}v)_t+\nabla_x\bullet ({\tilde \rho}v\otimes v)+
       \nabla_x p(\rho)=0,\\
      & (\rho_0(x)-\bar\rho, v_0(x))=0,\ for\ |x|\ge R,
      \endcases\tag2.15$$
where,  $R>0$, and $0<\bar\rho\in (\rho_*,\rho^*)$ 
satisfies the subluminal condition,

$$p'(\rho^*)\le c^2.\tag2.16$$
 
\proclaim{Lemma 2.3} Let $(\rho, v)(x,t)$ be a $C^1$ solution
of the Cauchy problem (2.15) (equivalent to (1.3) with the same initial data),
where the initial data $(\rho_0, v_0)(x)$ takes
values in
a compact subset of $\Omega_z$ (c.f. (2.3)). 
Then the support of $(\rho-{\bar \rho},v)(x,t)$ is contained in the ball 
$B(t)=\{x: |x|\le R+st\}$ where
$$s=\sqrt{p'(\bar\rho)}$$
is the sound speed in the far field.
\endproclaim\par

\demo{ Proof} This lemma is a consequence of the local
energy estimates. It will be proved using the method of [11] for
symmetric hyperbolic systems.  For this purpose, 
we first observe that $w=(\nabla_\theta \eta)^T$ given in (2.10)
(where $\eta$ is as in (2.9)) renders the system
(2.15) symmetric hyperbolic [3]:

$$A^0(w) \frac{\partial w}{\partial t} 
+\sum_{i=1}^{3} A^i(w)\frac{\partial w}{\partial x_i}=0,\tag2.17$$
where the coefficient matrices $A^\alpha(w)=(A_{mn}^\alpha)$, 
$\alpha, m, n=0, 1, 2, 3$ are given by (6.5); that is
$$ A^0=(\nabla_{\theta}^2 \eta)^{-1},\ 
   A^k=(\nabla_{\theta} f^k) (\nabla_{\theta}^2\eta)^{-1}.$$
We now compute the
explicit form of these matrices. First of all, we have
$$A^0=(\nabla_{\theta}^2\eta)^{-1}=\Psi(\rho)
 \left( \matrix a_1&a_2v^T\\
     a_2v&a_3vv^T+a_4I_3  
      \endmatrix\right),\tag2.18$$
where $I_3$ is the $3\times3$ identity matrix, and
 $$\Psi(\rho)=\frac{1}{K}(\rho c^2+p)^2 e^{-\phi(\rho)},\tag 2.19$$
     and
     $$\aligned & a_1=\frac{c^4+3p'v^2}{c^3p'(c^2-v^2)^{3/2}},
      \ a_2=\frac{c^4+2p'c^2+p'v^2}{c^3p'(c^2-v^2)^{3/2}},\\
     & a_3=\frac{c^2+3p'}{cp'(c^2-v^2)^{3/2}},
   \ a_4=\frac{1}{c(c^2-v^2)^{1/2}}.
    \endaligned\tag2.20$$
For $A^k$, $k=1, 2, 3$, we first compute 
$$\aligned (\nabla_\theta f^k)&=(\nabla_z f^k)(\nabla_z\theta)^{-1}\\
                     &=\left( \matrix 0& e_k^T\\
     [c^2(c^2+v^2)p'E_1e_k-c^2(c^2+p')E_1v_kv ] & [ -C_4ve_k^T+v_kI_3]  
      \endmatrix\right), \endaligned\tag2.21$$
with $e_k=(\delta_{1k}, \delta_{2k}, \delta_{3k})^T$, where
$(\nabla_z f^k)$ is given by
$$(\nabla_z f^k)= \left( \matrix B_3v_k&B_2v_kv^T+B_4 e_k^T\\
     B_3v_kv+p'e_k&  B_2v_kvv^T+B_4ve_k^T+B_4v_kI_3  
      \endmatrix\right);$$
c.f. (6.13) below. Then we have

$$\aligned A^k&=(\nabla_\theta f^k)(\nabla_{\theta}^2\eta)^{-1}
               =(\nabla_\theta f^k)A^{0}\\
    &=\Psi(\rho)\left( \matrix a_2v_k& [a_3v_kv^T+a_4 e_k^T]\\
    [ a_3v_kv+a_4e_k] & [ a_3v_kvv^T+a_4v_kI_3+a_4(e_kv^T+ve_k^T) ] 
      \endmatrix\right). \endaligned\tag2.22$$
It is clear that the matrices $A^k(w)$ are all 
real symmetric and smooth
in $\Omega_z$. Furthermore $A^0(w)=(\nabla^2 \eta)^{-1}$ is
positive definite in $\Omega_z$. \par

Now, we choose $\rho_m=\bar\rho$ subject to (2.16)  for convenience. 
In this setting, the background state in the $w$-variable becomes 
${\bar w}=w(\bar\rho,0)=0$.
Set
$${\tilde A}^i(w)=(A^0(w))^{-1}A^i(w),$$
and define
$$Q(\lambda,\xi)=
\lambda I_4-\sum_{\alpha=1}^3 \xi_\alpha {\tilde A}^\alpha(0), 
\tag 2.23$$
where $(\lambda, \xi)\in {\bold R}\times S^2$ ($S^2$ is the unit 2-sphere). 
Using the real symmetry of 
$ {A}^\alpha(0)$, for each $\xi\in S^2$, we see that 
the characteristic equation
$$det Q(\lambda, \xi)=0,$$
has real roots $\lambda_i(\xi)$, $i=0, 1, 2, 3$,  
called the characteristic speeds. 
Let $\bar\lambda$ be the largest absolute value of these characteristic 
speeds. For any fixed $(x_0, t_0)\in {\bold R}^3\times (0, T_\infty)$, 
we define the family of cones
$$C_\tau=\{(s, x):|x_0-x|\le {\bar\lambda}(t_0-s),\ 0\le s\le \tau\},
\tag 2.24$$
parametrized by $\tau\in [0, t_0)$ and the associated cross sections 

$$ E_\sigma=\{(\mu, x)\in C_\tau: \mu=\sigma\}.
        \tag 2.25$$
We introduce the linear partial differential operator
  $$P=A^0(0)\frac{\partial}{\partial t} 
 +\sum_{\alpha=1}^3 A^\alpha (0) \frac{\partial}{\partial {x_\alpha}},
\tag2.26$$
where again $0$ is the background state in the $w$-variable. 
The equation (2.17) reads

$$Pw=(A^0({0})-A^0(w))\frac{\partial w}{\partial t}
       +\sum_{\alpha=1}^3 (A^\alpha (0)-A^\alpha (w)) 
    \frac{\partial w}{\partial{x_\alpha}}.\tag2.27$$ 
Now we multiply both sides of (2.27) by $2w^T$,  to get

$$\aligned &\partial_t [w^TA^0(0)w]+\sum_{\alpha=1}^3 
  \frac{\partial}{\partial {x_\alpha}} [w^TA^\alpha (0) w]\\
  &=2w^T\left[(A^0({0})-A^0(w))\frac{\partial w}{\partial t}
       +\sum_{\alpha=1}^3 (A^\alpha (0)-A^\alpha (w)) 
    \frac{\partial w}{\partial {x_\alpha}}\right].
    \endaligned\tag 2.28$$
We integrate (2.28) over $C_\tau$ to obtain

$$\aligned &\int_{C_\tau}\sum_{\alpha=0}^3\partial_\alpha 
 [ w^T A^{\alpha}(0)w](x,\sigma)\ dxd\sigma \\
  &=\int_{C_\tau}2w^T\sum_{\alpha=0}^3
(A^\alpha(0)-A^\alpha(w))\partial_\alpha w\\
 &\le C\underset{C_\tau}\to{ \max}|\nabla w|
                    \int_0^\tau\int_{E_\sigma}  |w|^2(x,\sigma)\ dxd\sigma.
                    \endaligned\tag2.29$$
Where, $\partial_0=\partial_t$, and we have used the mean value 
theorem in the estimation of the last step. For the left hand side
of (2.29), we want to apply the divergence theorem since it is in
divergence form. For this purpose, we need to determine the boundary of
$C_\tau$ and the associated unit outer normal vector. The boundary of
$C_\tau$ consists of three parts: the cap $E_\tau$ with unit outer normal
$(1, 0,0,0)^T$, the base $E_0$ with unit outer normal $(-1,0,0,0)^T$,
and along the surface 
$$ R_\tau=\{(\sigma, x): |x_0-x|={\bar\lambda}(t_0-\sigma), 0\le \sigma\le\tau\}\tag 2.30$$
the unit outer normal vector is
$$ n= \frac{1}{\sqrt{1+{\bar\lambda}^2}}({\bar\lambda}, -\nu^T)^T,
\  \nu=\frac{(x_0-x)}{|x-x_0|}.\tag 2.31$$
To see (2.31), we note that on $R_\tau$ one has
$$ ({\bar\lambda}, -\nu^T)^T\bullet ((t_0-\sigma), (x_0-x)^T)^T=0.$$
We now apply the divergence theorem to the left hand side of (2.29),
$$\aligned &\int_{C_\tau}\sum_{\alpha=0}^3\partial_\alpha 
  [w^T A^{\alpha}(0)w](x,\tau)\ dxd\tau\\
&=\int_{E_\tau}(w^TA^0(0)w)(x, \tau)\ dx-\int_{E_0}(w^TA^0(0)w)(x, 0)\ dx\\
  &
+\frac{1}{\sqrt{{\bar\lambda}^2+1}} 
\int_0^\tau \int_{\partial E_\sigma}({\bar\lambda}w^TA^0(0)
                   w-w^T \sum_{\alpha=1}^3 
    \nu_\alpha
                    {A}^\alpha(0) w )(x, \sigma) \ dS_xd\sigma,
                    \endaligned\tag2.32$$
where $dS_x$ denotes the surface element on $\partial E_\sigma$. 
The third term of (2.32) on the right hand side can be simplified as follows:
$$\aligned & ({\bar\lambda}w^TA^0(0)
     w-w^T \sum_{\alpha=1}^3 \nu_\alpha
                    {A}^\alpha(0) w ) \\
                    &= w^T A^0(0) ({\bar\lambda}I_4-
                     \sum_{\alpha=1}^3 \nu_\alpha
                   {\tilde A}^\alpha(0)) w\\
         &= w^T A^0(0) Q({\bar\lambda}, \nu)w.
                   \endaligned\tag 2.33$$

We recall that $A^0(0)>0$ and

$$A^0(0)Q(\lambda,\nu)=\lambda A^0(0)-\sum_{\alpha=1}^3 \nu_\alpha A^\alpha(0)\tag 2.34$$
is real symmetric. We claim that for any $\nu\in S^2$,
$$A^0(0)Q({\bar\lambda},\nu)\ge 0,\tag 2.35$$
which will be verified at the end of this proof.

Therefore, we conclude from (2.29), (2.32), (2.33) and (2.35)  that
 $$\aligned&\int_{E_\tau}
(w^T A^0(0)w)(x, \tau)\ dx\\
                 & \le \int_{E_0}( w^T A^0(0)w)(x,0)\ dx
                 + C_1\underset{C_\tau}\to{ \max}|\nabla w|
                    \int_0^\tau\int_{E_\sigma}  |w|^2(x, \sigma)\ dxd\sigma.
                    \endaligned\tag 2.36$$
 Since $A^0(0)>0$, 
there are positive constants $C_2$ and $C_3$
 such that
  
  $$\int_{E_\tau} | w|^2(x, \tau)\ dx
\le C_2\int_{E_0}| w|^2(x,0)\ dx
   +C_3\int_0^\tau\int_{E_\sigma}| w|^2(x, \sigma)\ dxd\sigma,$$
  which, by Gronwall's inequality implies that 
  
  $$\int_{E_\tau}  | w|^2(x,\tau)\ dx
\le C_2e^{C_3\tau}(\int_{E_0} |w|^2(x,0)\ dx) .\tag 2.37$$
Therefore, if $w(x,0)=0$ for $|x-x_0|\le {\bar\lambda}t_0$, then 
  $w(x,\tau)=0$ for any $\tau\in [0, t_0)$ and 
    $|x-x_0|\le {\bar\lambda}(t_0-\tau)$.
  This implies that, if $w(x,0)=0$ for $|x|>R$, 
then $w(x,t)=0$ for $|x|>R+{\bar\lambda}t$.\par
  
The next step is to verify that ${\bar\lambda}=\sqrt{p'(\bar\rho)}$. 
For this purpose, we compute the largest possible characteristic speed at 
a constant background state.  Now we compute the eigenvalues of
$$\sum_{\alpha=1}^3\xi_\alpha{\tilde A}^\alpha(0),\ ~ ~ ~ ~ \xi\in S^2.
\tag 2.38$$
Since
$${\tilde A}^\alpha
=(\nabla_{\theta}^2 \eta) ( \nabla_{\theta} f^\alpha) (\nabla_{\theta}^2 \eta)^{-1},$$
 the matrix in (2.38) is similar to the matrix
 $$M(\xi)=\sum_{\alpha=1}^3 \xi_\alpha \nabla_{\theta}f^\alpha(\bar\theta),
\ \xi\in S^2,\tag2.39$$
 where ${\bar\theta}=(\bar\rho, 0, 0, 0)^T$ is the background state in the $\theta$-variable.  It is easy to
 compute:
 $$\aligned  \nabla_{\theta}f^\alpha(\bar\theta)&
 =\nabla_z f^\alpha (\nabla_z\theta)^{-1}|_{\rho=\bar\rho, v=0}\\
 &=\left( \matrix 0&e_\alpha^T\\
           p'(\bar\rho)e_\alpha&0 \endmatrix\right),\endaligned\tag 2.40$$
  where $e_i=(\delta_{1i}, \delta_{2i},\delta_{3i})^T$. Thus, one has
  $$M(\xi)= \left( \matrix 0&\xi^T\\
           p'(\bar\rho)\xi &0 \endmatrix\right).\tag2.41$$
 Now we claim that $M(\xi)$ has rank 2. This can be seen by 
 $$MM^T= \left( \matrix 1&0\\
           0 &p'(\bar\rho)^2\xi\xi^T \endmatrix\right),$$
 since $rank(\xi\xi^T)=1$. 
Thus, $M(\xi)$ has two non-zero eigenvalues and two zero
 eigenvalues. We need to find all non-zero eigenvalues of $M(\xi)$.
 We compute
 $$\aligned 0= \det(M(\xi)-r I_4)&=\det  \left( \matrix -r&\xi^T\\
           p'(\bar\rho)\xi &-rI_3 \endmatrix\right)\\
          & =\det  \left( \matrix 0&\xi^T\\
           (p'(\bar\rho)-r^2)\xi &-rI_3 \endmatrix\right)\\
           &=(p'(\bar\rho)-r^2)\det  \left( \matrix 0&\xi^T\\
           \xi &-rI_3 \endmatrix\right),\endaligned\tag 2.42$$
 and this implies $\pm \sqrt{p'(\bar\rho)}$ are the two distinct non-zero 
eigenvalues
 of $M(\xi)$. Therefore, we have
 $${\bar\lambda}=\sqrt{p'(\bar\rho)}.\tag2.43$$
Notice that we did not use (2.35) to obtain (2.43).

The last step is to verify (2.35). Since 
$\bar\lambda=s=\sqrt{p'(\bar\rho)}$, we have
$$ A^0(0)= \frac{K}{c^2} \left( \matrix \frac{1}{s^2}&0\\
           0 &I_3 \endmatrix\right),\ 
   A^\alpha(0)=\frac{K}{c^2} \left( \matrix 0&e_\alpha^T\\
           e_\alpha&0 \endmatrix\right),\ \alpha=1,2,3,$$
and thus
$$ A^0(0)Q({\bar\lambda}, \nu))= \frac{K}{c^2}
  \left(\matrix \frac{1}{s}&-\nu^T\\
           -\nu &sI_3 \endmatrix\right).\tag2.44$$
Let ${\bold r}=(r_0, r^T)^T$ be any 4-vector with $r\in {\bold R}^3$, 
we compute:
$$\aligned {\bold r}^T A^0(0)Q({\bar\lambda}, \nu)){\bold r}
          &=\frac{K}{c^2}(r_0, r^T) \left(\matrix \frac{1}{s}&-\nu^T\\
           -\nu &sI_3 \endmatrix\right)(r_0, r^T)^T\\
          &=\frac{K}{c^2}(\frac{1}{s}r_0^2-2r_0 \nu^T r+s|r|^2)\\
          &\ge \frac{K}{c^2}(\frac{1}{s}r_0^2-2r_0 |r|+s|r|^2)\\
          &= \frac{K}{sc^2} (r_0^2-2sr_0 |r|+s^2|r|^2)\\
          &=\frac{K}{sc^2}(r_0-s|r|)^2\ge 0.
          \endaligned\tag 2.45$$
Here, we have used
$$(\nu^T r)\le \sqrt{(\nu^2)(r^2)}=|r|.$$
Thus, we have proved (2.35).
The proof of this lemma is complete. \qquad\qquad\qed\enddemo\par

\subheading{3. Singularity Formation: Infinite Energy Case}\par

In this section, we prove the singularity formation of smooth solutions of 
(2.15) when the initial radial ``generalized" momentum is large. 
To begin, we prove the following two easy but useful identities.\par

\proclaim{ Lemma 3.1} ${\hat\rho}$ and ${\tilde \rho}$ satisfy the 
following identities:
$$\aligned
&\displaystyle
  {\hat\rho}=\frac{1}{c^2}{\tilde\rho} v^2+\rho,\\
   &\displaystyle {\tilde\rho}= \frac{1}{c^2}{\tilde\rho} v^2
                            + (\rho+\frac{p}{c^2}).
                            \endaligned\tag3.1$$
                            
  \endproclaim\par

\demo{Proof} From (2.1), it is easy to see

$$\aligned
\displaystyle {\hat \rho}
 &=\displaystyle [\frac{\rho c^2+p}{c^2-v^2}-\frac{p}{c^2}]\\
&=\displaystyle \frac{\rho c^4+pc^2-pc^2
   +pv^2}{c^2(c^2-v^2)}\\
&=\displaystyle \frac{\rho c^2c^2+pv^2}{c^2(c^2-v^2)}\\
&=\displaystyle \frac{\rho c^2v^2+\rho c^2(c^2-v^2)+pv^2}{c^2(c^2-v^2)}\\
&=\displaystyle \frac{\rho c^2+p}{c^2(c^2-v^2)} v^2
     +\frac{\rho c^2(c^2-v^2)}{c^2(c^2-v^2)}\\
&=\displaystyle \frac{1}{c^2}{\tilde \rho}v^2+\rho.
\endaligned$$

Hence, ${\tilde\rho}={\hat\rho}+\frac{p}{c^2}$ implies
$$ \displaystyle {\tilde \rho}
 =\frac{1}{c^2}{\tilde\rho} v^2
                            + \rho+\frac{p}{c^2}.
                            $$
                          \hskip10cm  \qquad\qquad\qed\enddemo\par

We again denote the sound speed in the far field by 
$$s=\sqrt{p'(\bar\rho)},$$
 and define the following quantities:
$$\aligned& \displaystyle
 M(t)=\int [{\hat \rho}(\rho, v)-{\hat\rho}({\bar\rho}, 0)](x,t) \ dx,\\
&\displaystyle
F(t)=\int {\tilde \rho}v\bullet x\ dx.\endaligned\tag3.2$$
By Lemma 2.3, both $M(t)$ and $F(t)$ are well-defined as long as the 
smooth solution exists. Using these two quantities,
we shall show that the smooth solution of (2.15) obtained in Theorem 2.1 blows up
in finite time if the initial data is subject to some restrictions. Roughly
speaking, if $M(0)>0$, and $F(0)>0$ is sufficiently large, then the solution will blow up in
finite time.  \par

\proclaim{ Theorem 3.2}  Assume that the initial data of 
(2.15) and $\bar\rho$ are chosen such that $M(0)>0$, $F(0)>0$ and 
$s^2<\frac13 c^2$. If
$$F(0)>\Gamma=\frac{32\pi s}{3(1-\frac{3s^2}{c^2})} R^4 \max {\hat\rho}_0(x),$$
then the smooth solution of the Cauchy problem (2.15) obtained in Theorem 2.1
blows up in finite time.
\endproclaim\par

\demo{Proof} Using Lemma 3.1, we know that

$$ \displaystyle {\hat\rho}(\rho, v)
  =\frac{1}{c^2} {\tilde \rho}v^2+\rho,$$
thus, 
$${\hat\rho}({\bar\rho}, 0)={\bar\rho}.
\tag3.3$$
This implies that
$$\displaystyle
\rho-{\bar\rho}= ({\hat\rho}(\rho, v)-{\hat\rho}(\bar\rho,0))
   -\frac{1}{c^2}{\tilde\rho}v^2.\tag3.4$$

From the first equation of (3.2), it is easy to see that
$$\aligned  \displaystyle \frac{d}{dt}M(t)&
       \displaystyle=\int ({\hat\rho}(\rho,v)-{\hat\rho}(\bar\rho,0))_t \ dx\\
                 & \displaystyle=\int {\hat\rho}_t\ dx\\
                  &\displaystyle=-\int \nabla_x\bullet ({\tilde \rho}v)\ dx\\
                  &=0,
                  \endaligned$$
where we have used the first equation in (2.2) or (2.15). Hence, 
$$  M(t)=M(0)=\int ({\hat\rho}_0(x)-{\bar\rho})\ dx>0.\tag3.5$$

Using the second equation in (2.2) and (3.2), we compute

$$\aligned
        F'(t)&=\displaystyle 
       \int ({\tilde \rho}v)_t \bullet x\ dx\\
   &=-\int [\nabla\bullet({\tilde \rho}v\otimes v)+\nabla   
    p(\rho)]\bullet x\ dx.\endaligned\tag3.6$$
 But if ${\bar p}=p(\bar\rho)$, we have
 
 $$\aligned \nabla p\bullet x&=\nabla (p-{\bar p})\bullet x\\
        &=\nabla\bullet[x(p-{\bar p})]-(p-\bar p)\nabla\bullet x\\
        &=\nabla\bullet[x(p-{\bar p})]-3(p-\bar p).\endaligned\tag 3.7$$
 We also note that 
        $$\aligned \nabla\bullet({\tilde \rho}v\otimes v)\bullet x
        &=\sum_{i,j=1}^3 \partial_{x_i}({\tilde \rho}v_iv_j )x_j\\
        &=\sum_{i,j=1}^3[ \partial_{x_i}({\tilde \rho}v_iv_j x_j)- {\tilde \rho}v_iv_j \partial_{x_i}x_j]\\
        &=\sum_{i,j=1}^3 [ \partial_{x_i}({\tilde \rho}v_iv_j x_j)- {\tilde \rho}v_iv_j \delta_{ij}]\\
        &=-{\tilde \rho}v^2+\sum_{j=1}^3 \nabla\bullet ({\tilde \rho} v\otimes v  x_j),\endaligned\tag 3.8$$
    where $v^2=v^Tv$.  Inserting (3.7) and (3.8) into (3.6), and 
using the divergence theorem, we obtain
        
        $$F'(t)=\int {\tilde p}v^2\ dx+3\int (p-{\bar p})\ dx.\tag 3.9$$
  
Since $p''(\rho)\ge 0$, $p'(\rho)$ is a non-decreasing function of $\rho$.
 It is clear that

$$p(\rho)-p(\bar\rho)=\int_{\bar\rho}^\rho  p'(\xi)\ d\xi \ge p'(\bar\rho)(\rho-\bar\rho).\tag3.10$$

Thus, using (3.4) and (3.5), one has
$$\aligned F'(t)&\ge \displaystyle\int {\tilde \rho} v^2\ dx +3s^2\int (\rho-{\bar\rho}) 
        \ dx\\
&=\displaystyle\int {\tilde \rho} v^2\ dx +3s^2M(t)
   -\frac{3s^2}{c^2}\int {\tilde\rho}v^2\ dx \\
&= \displaystyle (1-\frac{3s^2}{c^2})\int {\tilde \rho} v^2\ dx
                    +3s^2M(0)\\
&\ge \displaystyle (1-\frac{3s^2}{c^2})\int {\tilde \rho} v^2\ dx.
\endaligned\tag3.11$$

On the other hand,  we have the following estimate:

$$\aligned
F^2(t)&=\displaystyle (\int {\tilde \rho}v\bullet x\ dx)^2\\
       &\le \displaystyle (\int_{B(t)} |x|^2 {\tilde \rho}\ dx)
             (\int_{B(t)}{\tilde\rho}v^2\ dx)\\
        &\le \displaystyle 2 
            (\int_{B(t)} |x|^2 {\hat \rho}\ dx)
             (\int_{B(t)}{\tilde\rho}v^2\ dx),\endaligned\tag3.12$$
             where we have used the following fact:
             $${\tilde\rho}\le 2 {\hat \rho}.\tag3.13$$
             To see this, we note that from the subluminal condition
             $p'(\rho)<c^2$, together with $p(0)=0$, we get $p(\rho)\le c^2\rho$.
             Thus
             $${\tilde \rho}=\frac{1}{c^2}{\tilde \rho}v^2+\rho+\frac{p}{c^2}
                \le \frac{1}{c^2}{\tilde \rho}v^2+2\rho={\hat \rho}+\rho\le 2{\hat \rho}.$$
 Due to (3.3) and (3.5), we have the following estimate      
        
$$\aligned \displaystyle \int_{B(t)}{\hat \rho}|x|^2\ dx
        & \displaystyle \le (R+s t)^2\int_{B(t)} {\hat\rho}\ dx\\
       & \displaystyle = (R+s t)^2(M(t)+\int_{B(t)} \bar\rho\ dx)\\
      &= \displaystyle(R+s t)^2(M(0)+\int_{B(t)} \bar\rho\ dx)\\
       &= \displaystyle (R+s t)^2\int_{B(t)} {\hat\rho}_0(x)\ dx\\
       &\le \displaystyle \frac{4\pi}{3}(R+s t)^5
            (\max {\hat\rho_0(x)}).
\endaligned\tag 3.14$$
Hence, (3.12) gives
$$\aligned F^2(t) 
&\le \displaystyle \frac{8\pi}{3}(R+s t)^5
            (\max {\hat\rho_0(x)}) (\int_{B(t)}{\tilde\rho}v^2\ dx)\\
 &\displaystyle \equiv K_0 (R+s t)^5 (\int_{B(t)}{\tilde\rho}v^2\ dx),
\endaligned\tag3.15$$
where
$$K_0=\displaystyle \frac{8\pi}{3}
            (\max {\hat\rho_0(x)}).\tag3.16$$
Thus (3.11) and (3.15) imply that
$$\displaystyle
F'(t)\ge (1-\frac{3s^2}{c^2})K_0^{-1}(R+s t)^{-5} F^2(t),
\tag3.17$$
so
$$\displaystyle \frac{F'}{F^2}\ge K_1 (R+s t)^{-5},\tag3.18$$
where $K_1=(1-\frac{3s^2}{c^2})K_0^{-1}$.
Integrating (3.18) with respect to $t$, one has
$$\displaystyle
\frac{1}{F(t)}\le \displaystyle
\frac{1}{F(0)}-\frac{K_1}{4s}[R^{-4}-
         (R+s t)^{-4}]\equiv\psi(t).\tag3.19$$
Now $\psi(0)=\frac{1}{F(0)}>0$ by assumption, and

$$\psi(+\infty)=\frac{1}{F(0)}-\frac{K_1}{4s}R^{-4}<0,$$
if 
$$\displaystyle
  F(0)>\displaystyle \frac{4s R^4}{K_1}\equiv \Gamma.\tag3.20$$
Therefore, 
$$\frac{1}{F(t_0)}=0,\ for\ some\ t_0>0.\tag3.21$$ 
Thus the life-span $T$ 
of smooth solutions satisfies $T<t_0$. This completes the proof of Theorem 3.2.
\enddemo\hskip10cm\quad\qquad\qed\par

\subheading{4. Singularity Formation: Finite Energy Case}\par

Due to the hyperbolic nature of Einstein equations, one expects
the finite propagation speed of waves in the solutions. We will
prove in the following lemma that for any smooth solution with
compactly supported initial data, 
the support of the solution is invariant in time. 

\proclaim{Lemma 4.1} Let $(\rho, v)(x,t)$ be a smooth solution
of the Cauchy problem (1.3)--(1.4) up to some time $T>0$. 
If the support of initial data is contained 
in the ball $B_R(0)$ centered at the origin with radius $R$, 
then the support of $(\rho, v)(x,t)$ is contained
in the same ball $B_R(0)$ for any $t\in[0, T)$.
\endproclaim\par

\demo{Proof} Assume that the initial support of the solution is
contained in a ball $B_R(0)$,  the support of the smooth solution
will remain compact by the hyperbolic nature of the system (1.3). 
We denote by $x(t;x_0)$ the particle
path starting at $x_0$ when $t=0$, i.e.,

$$\frac{d}{dt}x(t; x_0)=v(x(t;x_0),t),\ \ \ x(t=0; x_0)=x_0,\tag 4.1$$
and by $S_p(t)$ the closed region that is the image of $B_R(0)$ 
under the flow map (4.1). Hence, the support of the smooth solution
of (1.3)--(1.4) will remain inside $S_p(t)$. Thus, 
fixing any $x_0$ on the boundary of $B_R(0)$, we have $\rho_0(x_0)=0$
and $v_0(x_0)=0$, and $x(t;x_0)$ is on the boundary of $S_p(t)$. 
Furthermore,
$$\frac{d}{dt}x(t; x_0)=v(x(t;x_0),t)=0,\tag 4.2$$
due to continuity of $v(x,t)$ and the fact that $x(t;x_0)$ sits
at the boundary of the support of the solution. 
Therefore, $x(t; x_0)=x_0$ for any $t\in[0, T)$ whenever $|x_0|=R$. Hence, $S_p(t)=B_R(0)$.
This proves this lemma. \enddemo\hskip10cm\qquad\qed\par

Based on Lemma 4.1, we shall prove the following blowup result.

\proclaim{Theorem 4.2} Suppose the support of the smooth 
functions $(\rho_0(x), v_0(x))$ is non-empty and contained 
in a ball $B_R(0)$ centered at the origin with radius $R$. 
Then the smooth solution of (1.3)-(1.4) with
the initial data $(\rho_0(x), v_0(x))$  blows up in finite time. 
\endproclaim\par

\demo{Proof}
We first introduce the following functions,
  $$ H(t)= \frac12 \int\ {\hat\rho}|x|^2\ dx,\ \
  F(t)= \int\ {\tilde\rho}v\bullet x \ dx,\ \
  E(t)=\int\ {\hat\rho}\ dx.\tag4.3$$
Here, $H(t)$ is the second moment of ${\hat \rho}$, $F(t)$ is the total radial ``generalized" momentum,
and $E(t)$ is the total ``generalized" energy.
These functions are well defined in the domain where 
the smooth solutions exist. 
Interesting relations between them can be
obtained by the following calculations. 

$E(t)$ is conserved, because using (2.2) one has

$$
 E'(t)=\int {\hat \rho}_t\ dx=-\int \nabla_x\bullet({\tilde \rho} v)\ dx=0.
$$
We thus have
$$ E(t)=E(0)=\displaystyle \int {\hat\rho}_0(x)\ dx>0,\tag4.4$$
for non-trivial initial data.

For $H(t)$, we have 

$$ \aligned
  H'(t)&=\displaystyle \frac12\int {\hat\rho}_t|x|^2\ dx\\
      &=\displaystyle -\frac12\int [\nabla_x\bullet ({\tilde \rho} v)] |x|^2\ dx\\
     &=\displaystyle \int {\tilde \rho}v\bullet x\ dx\\
     &=\displaystyle F(t),\endaligned\tag4.5$$
 where we have used the relation
  $$[\nabla_x\bullet({\tilde\rho}v)]|x|^2
   =\nabla_x\bullet({\tilde\rho}v|x|^2)-{\tilde\rho}v\bullet 2x.$$
From the second equation in (2.2) and integrating by parts, we have
$$ \aligned
  H''(t)=F'(t)&=\displaystyle \int ({\tilde \rho}v)_t \bullet x\ dx\\
   &=\displaystyle -\int ([\nabla_x\bullet({\tilde \rho}v\otimes v)]\bullet x
   +(\nabla_x p\bullet x))\ dx\\
           &=\displaystyle \int {\tilde \rho} v^2\ dx +\int 3p\ dx\\
         &=\displaystyle c^2\int_{B_R(0)} (\frac{1}{c^2}{\tilde \rho}v^2
           +\frac{3p}{c^2}) \ dx.\endaligned\tag4.6$$

By Jensen's inequality, we have
$$\aligned
 \displaystyle \int_{B_R(0)} p(\rho)\ dx
&=\displaystyle (\frac{4\pi}{3}R^3) 
   \frac{\int_{B_R(0)} p(\rho) \ dx}{(\frac{4\pi}{3}R^3)}\\
&\ge \displaystyle (\frac{4\pi}{3}R^3) 
     p(\frac{\int_{B_R(0)} \rho\ dx}{(\frac{4\pi}{3}R^3)}),
\endaligned\tag4.7$$
so (4.6) and (4.7) imply 
$$\aligned
  H''(t)&\ge \displaystyle c^2\left[\int_{B_R(0)} \frac{1}{c^2}{\tilde \rho}v^2 dx
+\frac{3}{c^2} (\frac{4\pi}{3}R^3) 
  p(\rho_B)\right]\\
 &\equiv  c^2 N(t),\endaligned\tag4.8$$
where 
$$\rho_B=\frac{\int_{B_R(0)} \rho\ dx}{(\frac{4\pi}{3}R^3)},$$
is the mean density over $B_R(0)$.
Since 

$$E(t)=E(0)=\displaystyle\int_{B_R(0)} (\frac{1}{c^2} {\tilde \rho}v^2
+\rho)\ dx,$$ 
it is possible to bound $N(t)$ from below using $E(0)$. 
We consider two cases. First if

$$\displaystyle\int_{B_R(0)} \frac{1}{c^2}{\tilde \rho}v^2 \ dx
\ge \frac12E(0),$$
we get
$$N(t)\ge \frac12E(0).\tag 4.9$$
On the other hand, if 
$$\displaystyle\int_{B_R(0)} \frac{1}{c^2}{\tilde \rho}v^2 \ dx
\le \frac12 E(0),$$
then as 
$$E(t)=E(0)=\displaystyle\int_{B_R(0)} (\frac{1}{c^2} {\tilde \rho}v^2
+\rho)\ dx\le \frac12 E(0)+\int_{B_R(0)}\rho \ dx,$$ 
we have 
$$\int_{B_R(0)} \rho\ dx\ge \frac12 E(0).$$ 
Thus
 $$\aligned N(t)&\ge\displaystyle \frac{3}{c^2} (\frac{4\pi}{3}R^3) 
  p(\rho_B)\\
&\ge \displaystyle\frac{3}{c^2} (\frac{4\pi}{3}R^3) 
  p(\frac12 E_B(0))\\
&\equiv B_1E(0)>0,\endaligned\tag4.10$$
where $B_1= \frac{4\pi R^3}{c^2 E(0)}
 p(\frac12 E_B(0)) $, and $E_B(0)=\frac{ E(0)}{(\frac{4\pi}{3}R^3)}$.
Define $B=c^2\min\{\frac12, B_1\}$; then (4.8)--(4.10) imply that
$$ H''(t)\ge BE(0)>0.\tag 4.11$$
This gives a lower bound on $H(t)$:

$$ H(t)\ge \frac12 BE(0)t^2+F(0) t+H(0).\tag 4.12$$

In order to refine (4.12), we estimate $F(t)$ in terms of $H(t)$ and $E(t)$.
Using (3.10), we have

$$\aligned 
|F(t)|&=\displaystyle|\int ({\tilde \rho}v\bullet x)\ dx|\\
     & \le\displaystyle (\int {\tilde \rho}|x|^2\ dx)^{\frac12}
        (\int {\tilde \rho}v^2\ dx)^{\frac12}\\
      &\le\displaystyle \sqrt{2}H(t)^{\frac12}(c^2E(t))^{\frac12}\\
&=\displaystyle c \sqrt{2}[H(t)E(t)]^{\frac12}\\
&\displaystyle\equiv D[H(t)E(t)]^{\frac12}.
\endaligned\tag4.13$$
We derive from (4.12) and (4.13) that

$$H(t)\ge \frac12 BE(0)t^2-D[H(0)E(0)]^{\frac12}t+H(0).\tag4.14$$

We note that (4.14) implies that $H(t)$ tends to 
infinity as $t$ goes to infinity. However,
we have the following uniform upper bound for $H(t)$:

$$ \aligned
   H(t)&=\frac12\int {\hat\rho}|x|^2\ dx\\
       &=\frac12\int_{B_R(0)} {\hat \rho}|x|^2\ dx\\
       &\le \frac12R^2\int {\hat \rho}\ dx\\
       &=\frac12E(0)R^2.\endaligned\tag4.15$$
Thus (4.12) or (4.14) together with (4.15) imply that the life-span of the smooth solutions must
be finite if $E(0)>0$. This completes the proof of Theorem 4.2.
\enddemo\hskip10cm\qquad\qed\par

Notice that, for non-trivial initial data, we have

$$\aligned&H(0)-\frac12E(0)R^2\\
               & =\frac12 \int {\hat\rho}_0|x|^2 \ dx
                 -\frac12 R^2\int {\hat\rho}_0\ dx\\
                &=\frac12 \int_{B_R(0)}{\hat\rho}_0(|x|^2-R^2)\ dx\\
                &<0.\endaligned\tag4.16$$
 This enables us to estimate the life-span as follows: from (4.14) and
(4.15) we have for smooth solutions 
  $$
    \frac12 BE(0)t^2-D\sqrt{H(0)E(0)} t
    +H(0)\le \frac12 E(0)R^2.$$
This is equivalent to 
$$\phi(t)= Bt^2-2Ddt+2d^2-R^2\le 0,\tag4.17$$
where $d^2=\frac{H(0)}{E(0)}$, and $\phi(0)<0$ by (4.16). 
Hence, the life-span $T$ of the smooth solution satisfies
$$
\displaystyle T\le \displaystyle\frac{{Dd}+\sqrt{D^2d^2-2Bd^2+R^2B}}{B}.
\tag 4.18$$

\subheading{5. Concluding Remarks}\par 

We have proved the blowup of smooth solutions of relativistic
Euler equations in both cases:  finite initial energy (Theorem 4.2)
and  infinite initial energy (Theorem 3.2).  In contrast to the characteristic 
method, we adapted the approach via some functions: total ``generalized" energy,
total radial  ``generalized" momentum, and the second moment. Our approach depends on the
beautiful structure of the equations and several quantities constructed from
the natural variables; 
c.f. Lemma 3.1. Although the relativistic Euler equations
are much more complicated than the classical Euler equations, 
these structures make our proofs possible.
We will now make some remarks on
our results and discuss some related issues.

\medskip

\proclaim{ Remark  1} In our blowup theorems, the velocity in the far field is 
assumed to be zero initially. For the more general case, say $v_0(x)={\bar v}$ off a bounded
set, the change of variables (Sideris {\rm  [11]})

    $$v\to v-{\bar v},\   x\to x+t{\bar v}$$
will reduce this problem to the case we considered. 
 
 \endproclaim\par
 
\proclaim{ Remark 2} The condition 
   $$p'(\bar\rho)< \frac{c^2}{3}$$ 
in Theorem 3.2 arises
naturally in the proof. Here, $3$ is the spatial dimension. In 
 $d$ dimensions, 3 is replaced by $d$. In particular, for $d=1$, this condition is 
that the sound speed is subluminal. For $p(\rho)=\sigma^2\rho$,
$d=1$, this condition guarantees the genuinely nonlinearity of the
relativistic Euler equations and allows the existence of global solutions
in BV; see Smoller and Temple {\rm [14]}.  This condition is not required
in Theorem 4.2.

\endproclaim\par

\proclaim{Remark 3}  Our blowup results crucially depend on
the compact support of the perturbations. Singularity formation
for more general initial data remains open. 

\endproclaim\par

\proclaim{Remark 4} The type of singularity which occurs is another 
open problem. The possibilities are: a) shock formation, 
 b) violation of the subluminal conditions; e.g. 
  $|v|$ tends to $c$, or $p'(\rho)\to c^2$,
c) concentration of the mass.

For $p(\rho)=\sigma^2\rho$ and $d=1$, the singularity must be a shock
if the initial data is away from the vacuum. It was shown in 
Smoller and Temple {\rm [14]} that
weak solutions exist globally in time with bounded total variation,
 subluminal velocity and positive density 
uniformly bounded from above and below. Furthermore, Smoller and Temple
proved in {\rm [14]} that the subluminal condition guarantees the
genuine nonlinearity of the equations, so one concludes that
the singularities in the solutions must be shocks by Lax' theory {\rm [4]}.
It would be interesting to clarify the types of singularities
for relativistic Euler equations in multi-dimensions.
 However, black hole formation is impossible for our problem, 
 since our spacetime is fixed to be flat  Minkowski  spacetime. 

\endproclaim\par

\proclaim{ Remark 5} The singularity in our Theorem 3.2 looks
like shock formation.  The largeness condition in radial ``generalized" momentum, 
 (3.20), implies that the 
   particle velocity must be supersonic in some region relative
   to the sound speed at infinity. One can guess that
    the singularity formation is detected
   as the disturbance overtaking the wave front thereby forcing the front 
   to propagate with supersonic speed. To see these things, 
 we argue as follows. 
    Using the fact
    ${\tilde \rho}\le 2{\hat\rho}$(c.f. (3.13)), one has
    
    $$F(0)\le  \frac{8\pi}{3} R^4(\max {\hat\rho}_0(x))
    \max |v_0(x)|,
    \tag5.1$$
   while
   
   $$\aligned
   {\Gamma}&=\displaystyle \frac{4s}{1-\frac{3s^2}{c^2}}
     \frac{8\pi}{3} R^4\max {\hat\rho}_0(x)\\
      &\ge \displaystyle 
      4s\frac{8\pi}{3} R^4(\max {\hat\rho}_0(x)).
      \endaligned\tag5.2$$
    Hence, $F(0)> {\Gamma}$ implies that
    
    $$\max |v_0(x)|\ge 4s.\tag5.3$$
This insures the initial particle velocity is supersonic in some
region. However, the rigorous proof of shock formation 
is still open.   \endproclaim\par

\proclaim{Remark 6}   
The lower bound of the initial radial ``generalized" momentum 
in (3.20) depends on the initial velocity through ${\hat \rho}$. 
This is different 
from the Newtonian case, where ${\hat\rho}$ is replaced by ${\rho}$
and so in the Newtonian case it does not depend on the velocity. 
On the other hand, the velocity has to be subluminal.
Therefore, we must show that the set of initial data required in 
Theorem 3.2 is non-empty.
 From (5.1)--(5.3),  we find the set is non-empty if
 $$c>\max |v_0(x)|\ge \frac{4s}{1-\frac{3s^2}{c^2}}.\tag 5.4$$
 A simple calculation shows that the necessary condition for $s$ to satisfy is
 $$s<(\frac{\sqrt{7}}{3}-\frac23)c.\tag5.5$$
 Since ${\tilde \rho}\ge {\hat\rho}$, 
$F(0)$ is of the same order as the upper bound in (5.1).
 Thus, if $s$ is chosen to be small (this can be done by choosing $\bar\rho$
 small), 
one can easily find  initial data satisfying the conditions
 required in Theorem 3.2.
 \endproclaim\par

\proclaim{Remark 7} The equation of state $p(\rho)$ satisfying
 (1.5) is quite general for isentropic 
fluids. It can be weakened by replacing $p''(\rho)\ge 0$  
with  $p'(\rho)$ is non-decreasing.
This includes the well-known $\gamma$-law, 
$p(\rho)=\sigma^2\rho^\gamma$, $\gamma\ge1$
as a particular case.  In fact, in the case of a $\gamma$-law, 
(3.13) can be refined, and thus
(3.20) can be replaced by a weaker condition, as we now show. 

When $\gamma=1$, $s=\sigma$.  
(3.13) is refined as ${\tilde\rho}<(1+\frac{\sigma^2}{c^2}){\hat\rho}
={\tilde\rho}+ \frac{\sigma^2}{c^4}{\tilde\rho}v^2$ by Lemma 3.1. 
Thus, (3.20)  can be weakened to:
$$F(0)>\Gamma_1=\displaystyle \frac{4\sigma}{1-\frac{3\sigma^2}{c^2}}(1+\frac{\sigma^2}{c^2})
     \frac{4\pi}{3} R^4\max {\hat\rho}_0(x).\tag5.6$$
  
  When $\gamma>1$, we observe that the subluminal condition 
  
    $$p'(\rho)=\gamma \sigma^2\rho^{\gamma-1}\le c^2$$ 
 implies that $\frac{p}{c^2}\le \frac{1}{\gamma}\rho$. Thus, 
 $$\aligned(1+\frac{1}{\gamma}){\hat\rho}
 &=\hat\rho+\frac{1}{\gamma}\rho +\frac{1}{\gamma c^2}{\tilde\rho}v^2\\
 &\ge \hat\rho+\frac{1}{\gamma}\rho\\
 &\ge \hat\rho+\frac{p}{c^2}={\tilde\rho}.\endaligned$$
  We can thus refine (3.13)  to ${\tilde\rho}<(1+\frac{1}{\gamma}){\hat\rho}$, 
  and then (3.20) is replaced by the
  following weaker condition:
  $$F(0)>\Gamma_2=\displaystyle \frac{4s}{1-\frac{3s^2}{c^2}}(1+\frac{1}{\gamma})
     \frac{4\pi}{3} R^4\max {\hat\rho}_0(x).\tag5.7$$  
\endproclaim\par

\subheading{6. Appendix}\par
For reader's convenience, we justify the construction 
of a strictly convex entropy function
 for (1.3) due to Makino and Ukai in [7], and we will also correct several errors.
To this end,
 we first record (2.6)--(2.7) here ,
 
 $$\theta_t+\sum_{k=1}^3 (f^k(\theta))_{x_k}=0,\tag6.1$$
 where $\theta=(\theta_0, \theta_1, \theta_2,\theta_3)^T$ and 
  $f^k(\theta)=(\theta_k, f_1^k, f_2^k, f_3^k)$
 are defined by
 
 $$\aligned & \theta_0={\hat \rho},\ \theta_j={\tilde\rho}v_j,\\
    &f_j^k={\tilde\rho}v_jv_k+p\delta_{jk},\ j=1,2,3.\endaligned\tag6.2$$
 
 The scalar function $\eta=\eta(\theta)$ is called an entropy function 
and scalar functions $q^k(\theta)$,
 $k=1, 2, 3$ are called entropy flux functions, if they satisfy:
 
 $$\nabla_{\theta}\eta(\theta) \nabla_{\theta}f^k(\theta)
=\nabla_{\theta} q^k(\theta).\tag 6.3$$
Since the the right hand side of (6.3) is a gradient of the function $q^k$,
the relevant integrability condition (c.f. [1], page 39) is
$$(\nabla_{\theta}^2\eta) (\nabla_{\theta} f^k)
 =(\nabla_{\theta} f^k)^T(\nabla^2_\theta \eta). \tag 6.4$$

If we find such an $\eta$ that is strictly convex, the change of variables 
$\theta\to w=(\nabla_{\theta} \eta)^T$ will render (6.1) into
the symmetric form (2.17); see [1], where 
$$ A^0=(\nabla_{\theta}^2 \eta)^{-1},\ 
   A^k=(\nabla_{\theta} f^k) (\nabla_{\theta}^2\eta)^{-1}.\tag 6.5$$
To see this, we apply chain rule:
$$\partial_{\alpha} \theta=(\nabla_\theta w)^{-1} \partial_{\alpha} w
                         =(\nabla_{\theta}^2\eta)^{-1}\partial_{\alpha} w.
\tag6.6$$
Substituting (6.6) into (6.1), we obtain
$$ (\nabla_{\theta}^2\eta)^{-1} w_t 
 +\sum_{k=1}^3 (\nabla_{\theta} f^k)(\nabla_{\theta}^2\eta)^{-1}w_{x_k}=0.$$
 $A^0$ is positive definite if and only if $\eta$ is strictly convex.
To verify the real symmetry of $A^k$, we use (6.4). Multiplying both sides
of (6.4) by $(\nabla_{\theta}^2\eta)^{-1}$ on the  left and right, we see
 $ A^k=(A^k)^T.$\par

We will solve (6.3) keeping the mechanical energy of classical 
Euler equations in mind. Thus, instead of $\theta$, 
we will use $z=(\rho, v_1, v_2, v_3)^T$ as independent variables. We compute:

$$\nabla_z\theta=\left( \matrix B_1&B_2v^T\\
                   B_3v& B_2vv^T+B_4I_3 \endmatrix\right) ,\tag6.7$$
where
   $$\aligned &B_1=\frac{c^2+p'}{c^2-v^2}-\frac{p'}{c^2}, 
\ \  B_2=2\frac{\rho c^2+p}{(c^2-v^2)^2},\\
       &B_3=\frac{c^2+p'}{c^2-v^2},
\ \  B_4=\frac{\rho c^2+p}{c^2-v^2}.\endaligned.\tag 6.8$$
Moreover,                                                           
       $$ det(\nabla_z\theta)
=\frac{(\rho c^2+p)^3(c^4-v^2p')}{c^2(c^2-v^2)^4}>0,\tag6.9 $$
in $\Omega_z$.  
We can thus compute the inverse of $\nabla_z\theta$:
       
 $$(\nabla_z\theta)^{-1}=\left( \matrix c^2(c^2+v^2)E_1&-2c^2E_1v^T\\
                 -c^2(c^2+p')E_1E_2v& 2p'E_1E_2vv^T+E_2I_3 \endmatrix \right),
                \tag6.10$$
    with
    
   $$E_1=\frac{1}{c^4-p'v^2},\ \ \  E_2=\frac{c^2-v^2}{\rho c^2+p}.\tag6.11$$
   Based on (6.10), we will solve (6.3) using $z$ as 
independent variables for convenience. In the $z$-variables, (6.3) can becomes
   $$\nabla_z \eta C^k=D_z q^k,\ k=1,2,3,\tag 6.12$$
   where
$$(\nabla_z f^k)= \left( \matrix B_3v_k&B_2v_kv^T+B_4 e_k^T\\
     B_3v_kv+p'e_k&  B_2v_kvv^T+B_4ve_k^T+B_4v_kI_3  
      \endmatrix\right),\tag6.13$$
and
   $$\aligned C^k&=(\nabla_z\theta)^{-1}(\nabla_z f^k)\\
  & =\left( \matrix c^2C_1v_k&C_3 e_k^T\\
     -C_1C_2v_kv+C_2e_k& -C_4ve_k^T+v_kI_3  
      \endmatrix\right), \endaligned\tag6.14$$
   with
 $$\aligned& C_1=\frac{c^2-p'}{c^4-p'v^2},\ \ 
\  C_2=p'E_2=\frac{p'(c^2-v^2)}{\rho c^2+p},\\
   &C_3=\frac{c^2(\rho c^2+p)}{c^4-v^2p'},
\ \ \ C_4=\frac{p'(c^2-v^2)}{c^4-p'v^2}.\endaligned\tag 6.15$$
   Formally, (6.12) is an over-determined system, 
consisting of 12 equations for 4 unknowns.  We seek 
   solutions with the special form:
   $$\eta=\eta(\rho, y),\ q^k=Q(\rho, y)v_k,\ 
y=v^2=v_1^2+v_2^2+v_3^2.\tag 6.16$$
    to reduce the number of equations in (6.12).  
Substituting this ansatz into (6.12), we obtain
    the following first order linear system:
    $$\cases &\eta_y=Q_y,\\
                     &c^2C_1\eta_\rho+2C_2(1-C_1y)\eta_y=Q_\rho\\
                     &C_3\eta_\rho-2C_4y\eta_y=Q.\endcases\tag6.17$$
This seems still an over-determined system. However, it is possible
to derive a decoupled equation for $Q$ from (6.17). We first multiply
the second equation of (6.17) by $(\rho^2 c^2+p)$, 
and using $(c^2-p')C_3=c^2C_1(\rho^2 c^2+p)$, we have

$$(c^2-p')C_3\eta_\rho+2C_2(\rho c^2+p)(1-C_1y)\eta_y
=(\rho c^2+p)Q_\rho.\tag 6.18$$
Then, we compute $(c^2-p')\times(6.17)_3$:

$$(c^2-p')C_3\eta_\rho-2(c^2-p')C_4y\eta_y=(c^2-p')Q.\tag6.19$$

We subtract (6.19) from (6.18) and substitute $\eta_y$ with $Q_y$, using
(6.15); this reduces (6.17) into the following decoupled system:
     $$\cases &\eta_y=Q_y,\\
              &2(c^2-y)p'Q_y=(\rho c^2+p)Q_\rho- (c^2-p')Q.\endcases\tag6.20$$

    We now proceed to solve (6.20) with the help of (6.17). 
 First, $ (6.20)_1$ gives
    $$\eta=Q(\rho, y)+G(\rho).\tag 6.21$$
    Substitute this into $(6.17)_3$ to get
    $$G_\rho=\frac{c^2-y^2}{\rho c^2+p} Q-\frac{c^2-y^2}{c^2} Q_\rho, $$
    or equivalently 
    $$G_\rho=\frac{1}{\rho c^2+p} q-\frac{1}{c^2}q_\rho,\  \ q=(c^2-y)Q.
     \tag 6.22$$
    We observe that $G$ depends on $\rho$ only, 
so we have a linear first order
    ODE for $q$,
    $$\frac{1}{\rho c^2+p} q-\frac{1}{c^2}q_\rho=\frac{f(\rho)}{c^2},$$
    which has the solution
    $$q(\rho)=e^{\phi(\rho)}(g(\rho)+h(y)).\tag 6.23$$
    Here $\phi(\rho)$ is defined in (2.8).    
   Substituting (6.23) into $(6.20)_2$, 
    and separating  variables, one has 
    $$\frac{\rho c^2+p}{p'}\frac{dg}{d\rho}-g
=2(c^2-y)\frac{dh}{dy}+h=m(\rho,y).
 \tag 6.24$$
Where, the first term in (6.24) is independent of $y$, while the second term
 is independent of $\rho$. Thus, $m$ is independent of both $\rho$ and $y$. 
We conclude that $m=const.$. Thus, by integrating (6.24), we have
    $$\aligned & q=D_1(\rho c^2+p)+D_2e^{\phi(\rho)}\sqrt{c^2-v^2},\\
      &G=-\frac{D_1}{c^2} p+D_3,\\
      &Q=D_1\frac{\rho c^2+p}{c^2-v^2}
      +D_2\frac{e^{\phi(\rho)}}{\sqrt{c^2-v^2}},\\
      &\eta=D_1\frac{\rho c^2+p}{c^2-v^2}
        +D_2\frac{e^{\phi(\rho)}}{\sqrt{c^2-v^2}}
      -\frac{D_1}{c^2} p+D_3, \endaligned\tag6.25$$
      where $D_1$, $D_2$ and $D_3$ are integration constants. 
     With $K$ as in (2.8), one choice is
      $$D_1=c^2,\ D_2=-cK,\ D_3=0,$$
thus, 
  $$\eta=c^2{\hat\rho}-\frac{cKe^{\phi(\rho)}}{\sqrt{c^2-v^2}}.\tag6.26$$ 
The associated entropy-flux is $(q^1, q^2, q^3)^T$ defined by
$$q^k=\frac{c^2(\rho c^2+p)}{c^2-v^2}v_k
      -\frac{cKe^{\phi(\rho)}}{\sqrt{c^2-v^2}}v_k.\tag6.27$$
Moreover $\eta$ is strictly convex as was verified in Section 2.\par

\refstyle{C}

\enddocument
\bye